\documentclass[multphys,vecphys]{svmult}

\usepackage{makeidx}         
\usepackage{graphicx}        
\usepackage{multicol}        
\usepackage[bottom]{footmisc}

\makeindex             

\begin{document}

\title*{Evolution of supermassive black holes}
\author{Marta Volonteri\inst{1}}
\institute{Institute of Astronomy, University of Cambridge, Madingley Road, Cambridge CB3 0HA, UK
\texttt{marta@ast.cam.ac.uk}}
%
%
\maketitle
\begin{abstract}
Supermassive black holes (SMBHs) are nowadays believed to reside in most local galaxies, and the 
available data show an empirical correlation between bulge luminosity - 
or stellar velocity dispersion - and black hole mass, suggesting a single mechanism for 
assembling black holes and forming spheroids in galaxy halos. The evidence
is therefore in favour of a co-evolution between galaxies, black holes and quasars.
In cold dark matter cosmogonies, small-mass subgalactic systems 
form first to merge later into larger and larger structures. In this paradigm galaxy halos experience multiple
mergers during their lifetime. If every galaxy with a bulge hosts a SMBH in its center, 
and a local galaxy has been made up by multiple mergers, then a black hole binary is a natural 
evolutionary stage. The evolution of the supermassive black hole population clearly has to be investigated 
taking into account both the cosmological framework and the dynamical evolution of SMBHs and their hosts. 
The seeds of SMBHs have to be looked for in the early Universe, as very luminous quasars are 
detected up to redshift higher than $z=6$. These black holes evolve then in a hierarchical fashion,
following the merger hierarchy of their host halos. Accretion of gas, traced by quasar activity,
plays a fundamental role in determining the two parameters defining a black hole: mass and spin. 
A particularly intriguing epoch is the initial phase of SMBH growth. It is very challenging to meet 
the observational constraints at $z=6$ if BHs are not fed at very high rates in their infancy. 
\end{abstract}

\section{Introduction}
\label{sec:1}

It is well established observationally that the centers of most galaxies host supermassive black holes (SMBHs) with masses in the
range $M_{BH} \sim 10^6-10^9\,M_\odot$ (e.g., Ferrarese et al. 2001; Kormendy \& Gebhardt 2001; Richstone 2004).  The
evidence is particularly compelling for our own galaxy, hosting a central SMBH with mass $\simeq 4\times10^6\,M_\odot$  
(Ghez et al. 2000, 2005; Sch\"odel et al. 2002)). Dynamical estimates indicate that, across a wide range, the central
black hole (BH) mass is about 0.1\% of the spheroidal component of the host galaxy (H\"aring \& Rix 2004; Marconi \& Hunt 2004; 
Magorrian et al. 1998).  A tight correlation is also observed between the BH mass and the stellar velocity dispersion of the hot stellar 
component (Ferrarese \& Merritt 2000; Gebhardt et al. 2000; Tremaine et al. 2002).  
For galaxies, these correlations may well extend down to the smallest masses.  For example, the dwarf Seyfert~1 galaxy POX 
52 is thought to contain a BH of mass $M_{BH} \sim 10^5\,M_\odot$ (Barth et al. 2004). At the other end, however, the Sloan Digital 
Sky survey detected luminous quasars at very high redshift, $z>6$. Follow-up observations confirmed that at least some of 
these quasars are powered by supermassive black holes (SMBHs) with masses $\simeq 10^9\, M_\odot$ (Barth et al. 2003; Willott et al. 2003). 
We are therefore left with the task of explaining the presence of very big SMBHs when the Universe is less than {\rm 1 Gyr} old, 
and of much smaller BHs lurking in {\rm 13 Gyr} old galaxies.

\section{Sowing the seeds of black holes}
SMBHs with $M_{BH} \sim 10^6-10^9\,M_\odot$ cannot form from simple stellar evolution of comparable mass stars. They probably evolve from
seeds of intermediate mass. One first possibility is the direct formation of a BH from a collapsing gas cloud (Haehnelt \& Rees 1993; 
Loeb \& Rasio 1994; Eisenstein \& Loeb 1995; Bromm \& Loeb 2003; Koushiappas, Bullock \& Dekel 2004). The main issue for this family 
of models is how to get rid of the angular momentum of the gas, so that it can form a central massive object, which eventually becomes subject 
to post-Newtonian gravitational instability and forms the seed BH. The mass of the seeds vary according to different models, but typically are 
in the range $M_{BH} \sim 10^3-10^6\,M_\odot$. An alternative scenario proposes the gravitational core collapse of relativistic star clusters 
which may have been produced in starbursts at early times (Shapiro 2004). This scenario leads to a formation of seed BHs in the mass range
$M_{BH} \sim 10^2-10^4\,M_\odot$. The seeds of SMBHs can also be associated with the remnants of the first generation of stars. The first
stars which form at $z\sim 20$ in halos which represent high-$\sigma$ peaks of the primordial density field. At zero metallicity, 
the gas cooling proceeds in a very different way compared to local molecular clouds. The inefficient cooling might lead 
to a very top-heavy initial stellar mass function, and in particular to the production of very massive stars with masses $>100 M_\odot$ 
(Carr, Bond, \& Arnett 1984). If very massive stars form above 260 $M_\odot$,  they would rapidly collapse to massive BHs
with little mass loss (Fryer, Woosley, \& Heger 2001), i.e., leaving behind seed BHs with masses $M_{BH} \sim 10^2-10^3\,M_\odot$ 
(Madau \& Rees 2001).  

It is not possible to distinguish among the proposed models with current observations: the initial conditions are mostly washed out 
when SMBHs gain most of their mass, between $z=3$ and $z=1$.
There are two ways we can find out in which mass range the seed BHs were formed. One is to detect a subpopulation
of relic seed BHs still lurking in present-day galaxy halos. The prediction of a population of massive BHs wandering 
in galaxy halos and the intergalactic medium at the present epoch is an inevitable outcome, if the assembly history of SMBHs 
goes back to early times. These intermediate mass BHs have been left over by inefficient galaxy mergers, or dynamical interactions
(see \S3). The number and mass function of these wandering BHs differs largely from model to model. If these relic seeds are 
massive, $M_{BH} \sim 10^5-10^6\,M_\odot$, they can perturb the gravitational potential, and generate wave patterns in the gas within 
a disc close to galaxy centers. The observed density pattern can be used as a signature in detecting the most massive
wandering black holes in quiescent galaxies (Etherington \& Maciejewski 2006). Several attempts have been made to 
investigate the emission properties of these wandering BHs (Islam, Taylor \& Silk, Volonteri \& Perna 2005, Mii \& Totani 2005), and 
their possible association to the population of ultraluminous off-nuclear (`non-AGN') X-ray sources (ULXs) that have been detected 
in nearby galaxies (van der Marel 2003, Fabbiano 2005). No strong conclusion has been reached yet. 

\begin{figure}
\centering
\includegraphics[height=7cm]{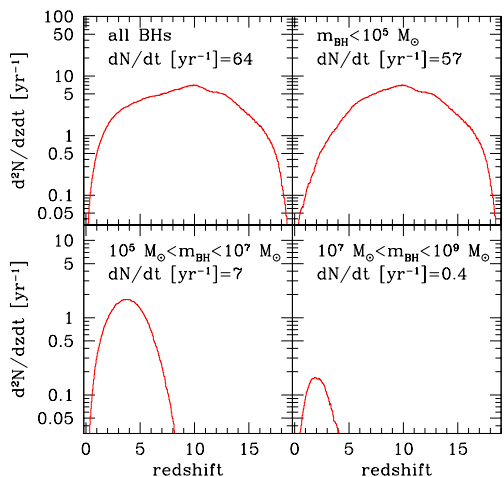}
\caption{Predictions on the number of MBH binary coalescences observed by LISA per year at $z=0$, 
per unit redshift, in different BH mass intervals. Each panel also lists the
integrated event rate, $dN/dt$, predicted by Sesana et al. (2004). If SMBH form from very massive seeds at low 
redshift, the expected merger rate is of order a few per year (cfr. bottom panel). If SMBHs evolve from high redshift
small seeds the merger rate can be up to tens or hundred per year (cfr. upper panels). }
\label{mvfig:1}       
\end{figure}

A second way to discern the seeds of SMBHs is through gravitational waves. Merging black holes are expected to 
be a strong source of gravitational waves, and the planned gravitational wave interferometer LISA is expected to
detect black hole mergers out to very large redshifts, and down to very small masses, being sensitive to mergers
of BHs with masses in the range $10^3-10^7M_\odot$. The difference in prediction between models can be up to 2-3 orders of
magnitude (Haehnelt 2004), and therefore LISA will be extremely helpful in solving this riddle (Figure \ref{mvfig:1}).

\section{Spinning top toys and yo-yos}
\begin{figure}
\centering
\includegraphics[height=7cm]{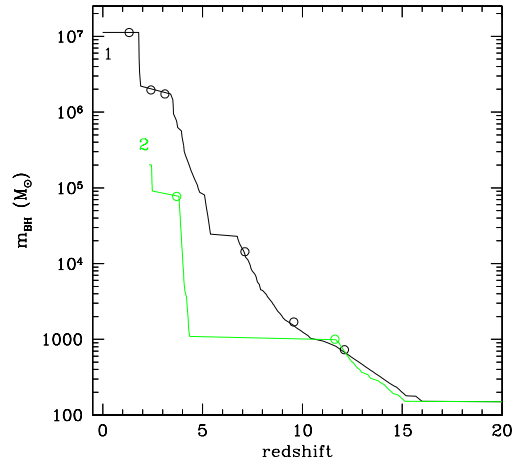}
\caption{Mass-growth history of two SMBHs, one ending in a massive 
halo (`1') at $z=0$, and one in a satellite (`2') at $z=2.3$. 
The SMBH mass grows by gas accretion after major merger events and by SMBHs mergers ({\it circles}).  
Note how most of the mass of the black holes is gained in accretion episodes and not through SMBH mergers.}
\label{mvfig:2}       
\end{figure}
Whatever your favourite choice of seed holes, several models, starting from different seeds, have proved 
equally successful in explaining the evolution of SMBHs and quasars in the redshift range $1<z<3$ 
(Kauffmann \& Haehnelt 2000; Wyithe \& Loeb 2002, 2003 Cattaneo et al. 1999). 
Accretion episodes are believed to be triggered by major mergers of galaxies, which can destabilize the gas 
in the interacting galaxies. The cold gas may be eventually driven into the very inner regions, 
fueling an accretion episode and the growth of the nuclear SMBH (Fig. \ref{mvfig:2}).

This last year has been especially exciting, as the first high resolution simulations of 
galactic mergers including BHs and quasar feedback, showed that the co-evolution scenario 
is generally correct (Springel, di Matteo \& Hernquist 2005;  di Matteo, Springel \& Hernquist 2005). 
The complete and detailed picture, in particular regarding the physics in the vicinity of the SMBH, 
is not clear yet, as simulations do not allow one to resolve the sphere of influence of the BHs. 

The above mentioned models for the evolution of SMBHs and quasars all reproduce satisfactorily the observed luminosity 
function of quasars. Integration, over redshift and luminosity, of the luminosity function of quasars, gives the total light emitted by 
known quasars (Yu \& Tremaine 2002; Elvis, Risaliti, \& Zamorani 2002; Marconi et  al. 2004). Using the Soltan's argument, 
this quantity can be converted 
into the total mass density accreted by black holes during the active phase, by assuming a mass-to-energy conversion efficiency, 
$\epsilon$, and normalizing to the local black hole mass  density (Aller \& Richstone 2002; Merloni, Rudnick \& Di Matteo 2004). 
The average radiative efficiency results somehow larger than $\epsilon=0.06$, which is the standard value for non-spinning BHs. 

The spin of a SMBH is modified by a merger with a secondary BH, or by interplay with the accretion disc and magnetic fields. 
If accretion proceeds from a thin disk, and magnetic processes are not important, rapidly spinning SMBHs are to be expected 
(Volonteri et al. 2005). BH mergers, in fact, do not lead to a systematic effect, but simply cause the  BHs to random-walk 
around the spin parameter they had at birth.  BH spins, instead, efficiently couple with the angular momentum of the accretion 
disk, producing Kerr holes independent of the initial spin. This is because, for a thin accretion disk, the BH aligns  
with the outer disk on a timescale that is much shorter than the Salpeter time, corresponding to an e-folding time for 
accretion at the Eddington rate, leading to most of the accretion being from prograde equatorial orbits. 
In any model in which SMBH growth is triggered by major mergers, every accretion episode must typically double the BH mass 
to account for the local SMBH mass density. 
As a result, most of the mass accreted by the hole acts to spin it up, even if the orientation of the spin axis 
changes in time. Most individual accretion episodes thus produce rapidly-spinning BHs independent of the initial spin.

\begin{figure}
\centering
\includegraphics[height=7cm]{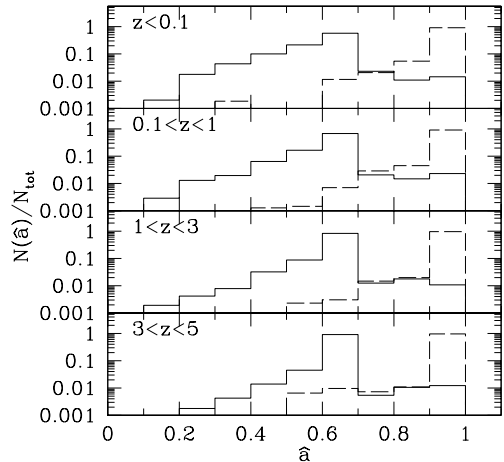}
\caption{Distribution of SMBH spins in different redshift intervals. ``Seed'' holes are born with a spin 
parameter $\hat a=0.6$.
{\it Solid  histogram:} effect of black hole binary coalescences only. 
{\it Dashed histogram:} spin distribution from binary coalescences and gas accretion via a thin disk.
}
\label{mvfig:3}       
\end{figure}

Our investigation finds that the spin distribution is skewed towards rapidly spinning holes,
is already in place at early epochs, and does not change significantly below
redshift 5. As shown in Fig. \ref{mvfig:3}, about 70\% of all SMBHs have spins larger than $\hat a≡ a/m_{BH} = 0.9$, 
corresponding to efficiencies approaching 20-30\% (assuming a ``standard'' spin-efficiency conversion).
The spin parameter of a few SMBHs in the local Universe seems indeed to be non-zero, as shown by
the prominent soft X-ray excess (Crummy et al. 2006), or the profile of the Iron line (Miniutti, Fabian \& Miller 2004, 
Streblyanska et al. 2005). 

Highly spinning black holes pose a challenge at early times, however. 
The accretion of mass at the Eddington rate causes the BH
mass to increase in time as
\begin{equation}
M(t)=M(0)\;\exp\left(\frac{1-\epsilon}{\epsilon}\frac{t}{t_{\rm Salp}}\right),
\end{equation}
where $t_{\rm Salp}=0.45\, \epsilon/(1-\epsilon)\,{\rm Gyr} $. Given an initial BH mass, $M(0)$, the higher the 
efficiency, the longer it takes for the BH to grow in mass up to its final mass (Shapiro 2005).
The highest redshift quasar currently known, SDSS 1148+3251, at $z=6.4$,
has estimates of the SMBH mass in the range $(2-6)\times 10^9 M_\odot$ (Barth et al. 2003, 
Willott, McLure \& Jarvis 2003).  If the accretion efficiency is 40\%, and the SMBH powering SDSS 1148+3251
has been accreting continuously for the whole Hubble time (about $1 {\rm Gyr}$ at $z=6$), then it has grown 
by a factor $\simeq 1000$, requiring a $10^6M_\odot$ seed well before halos of comparable mass have populated the Universe. 

A way out is to consider a large contribution to the BH mass by mergers. 
At $z<5$ MBH mergers do not play a fundamental role in building up the mass of SMBHs
(Yu \& Tremaine 2003), but they can be possibly important at $z>5$, where we do not 
have constraints from a Soltan-type argument. Mergers can have a positive contribution
to the build-up of the SMBH population that has been suggested could be up to $10^9 M_\odot$
(Yoo \& Miralda-Escud\'e 2004). On the other hand, dynamical processes can disturb the 
growth of BHs, especially at high redshift (Haiman 2004, Yoo \& Miralda-Escud\'e 2004), 
and give a negative contribution to SMBH growth.

\section{Playing pools with black holes}
The lifetime of BH binaries can be long enough (Begelman, Blandford \& Rees; 
Quinlan \& Hernquist 1997; Milosavljevic \& Merritt 2001; 
Yu 2002) that following another galactic merger a third BH can fall in and disturb the 
evolution of the central system.
The three BHs are likely to undergo a complicated resonance scattering interaction, leading to 
the final expulsion of one of the three bodies (`gravitational slingshot') and to the recoil of 
the binary. Any slingshot, in addition, modifies the binding energy of the binary, typically
creating more tightly bound systems.

Another interesting gravitational interaction between black holes happens
during the last stage of coalescence, when the leading physical 
process for the binary evolution becomes the emission of gravitational waves. 
If the system is not symmetric (e.g. BHs have unequal masses or spins) there would be a recoil due to 
the non-zero net linear momentum carried away by gravitational waves in the coalescence 
(`gravitational rocket'). The coalesced binary would then be displaced from the galactic 
centre, leaving straightaway the host halo or sinking back to the centre owing to dynamical 
friction. 

Slingshots and rockets basically give BHs a recoil velocity, that can even exceed the escape 
velocity from the host halo and spread BHs outside galactic nuclei (Volonteri, Haardt \& Madau 2003). 
In the shallow potential wells of mini-halos, the growth of BHs from seeds can be halted by these
ejections. Yoo \& Miralda-Escud\'e (2004) explored the minimum conditions that would allow the growth
of seed MBHs up to the limits imposed by SDSS 1148+3251 when recoils are taken into account. 
To meet the $z=6$ constraint, continued Eddington-limited accretion must occur onto MBHs forming 
in halos with $T_{\rm vir} >2000$K at $z\leq 40$. 

\begin{figure}
\centering
\includegraphics[height=7cm]{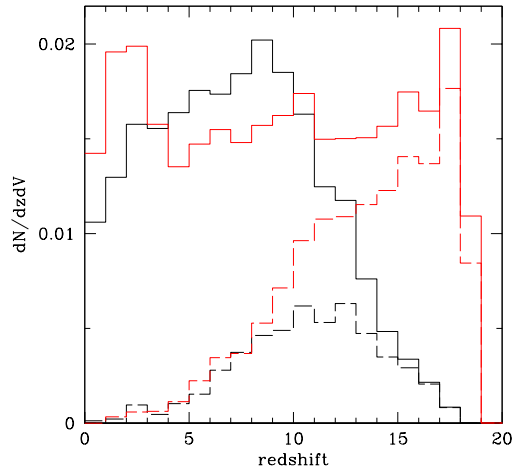}
\caption{Rates of binary MBH mergers ({\it solid line}) and ejections due to the 
gravitational radiation recoil ({\it dashed line}) per comoving Mpc$^3$. The 
subset of ejected binaries is selected requiring that the recoil velocity is larger than 
the escape velocity from the host halo. At redshifts $z>10$ about 80\% of the merging MBH 
binaries are ejected, and therefore lost, into the intergalactic medium. The black lines refer 
to a model with inefficient BH mergers, as binaries shrink by stellar dynamics. The red lines 
show the case in which BH binaries evolve on the viscous timescale, if an accretion disc is present.}
\label{mvfig:4}       
\end{figure}

One important piece of information which is still missing, is the typical timescale for a BH binary
to merge. After dynamical friction ceases to be efficient, at parsec scale or so, there is still 
a large gap before emission of gravitational waves can be efficient in bringing the binary to coalescence 
in less than a Hubble time. In massive galaxies at low redshift, the subsequent evolution of the 
binary may be largely determined by three-body interactions with background stars
(Begelman, Blandford, \& Rees 1980). In gas-rich, star-poor high redshift systems, gas processes
can lead the binary evolution. Dynamical friction in a gaseous environment is much more efficient than
against a stellar background, and, also, if an accretion disc is surrounding the MBH binary, the 
secondary BH can be dragged to the primary within a viscous timescale, much shorter than any other 
involved timescale (Armitage \& Natarajan 2005). Figure \ref{mvfig:4} compares the MBH merger rates in 
two different models accounting for the orbital evolution of MBH binaries in the phase 
preceding emission of gravitational waves. If gas processes (viscous drag, as shown) drive the 
MBH orbital decay, MBHs start merging efficiently at very early times, when host DM halos 
are still small. Though the absolute number of mergers is larger, more MBHs are therefore
ejected from DM halos due to the rocket effect, thus endangering the growth of SMBHs at high redshift. 
Volonteri \& Rees (2006) estimate that up to $50-80\%$ of black holes merging in high-redshift halos can be 
ejected due to the gravitational rocket effect.

\section{The origin of the largest SMBHs by means of natural selection}
As BH mergers in the early Universe can give a negative contribution to the growth of the SMBH population, 
we can turn back to the accretion properties of MBHs, and investigate them in more details. 

Metal--free halos with virial temperatures $T_{\rm vir} > 10^4$K can cool in
the absence of ${\rm H_2}$ via neutral hydrogen atomic lines to $\sim 8000$ K. 
The baryons can therefore collapse and settle into a rotationally supported dense disc at the center of the halo 
(Mo, Mao \& White 1998, Oh \& Haiman 2002). Oh \& Haiman showed that if only atomic line cooling operates, the 
discs are stable to fragmentation in the majority of cases, and this disc contains essentially all 
the baryons. In the rare cases  where fragmentation is possible, there would presumably then 
be star formation. However it seems inevitable that a substantial fraction of the baryons 
will remain in the fat disc at $5000{\rm K}<T_{\rm gas}<10^4{\rm K}$. 
This gas can supply fuel for accretion onto a MBH within it.  

If the rotationally supported disc is stable to bar instabilities, and therefore rotation in the gas disc is small. 
The infall of gas onto the central MBH is quasi radial and a tiny accretion disc forms. 
Estimating the mass accreted by the MBH within the Bondi-Hoyle formalism, the accretion rate is initially 
super-critical by a factor of 10 and grows up to a factor of about $10^4$ (Volonteri \& Rees 2005).
The size of the accretion disc which forms is of order of the trapping radius, i.e. the radius at which radiation is trapped as the infall 
speed of the gas is larger than the diffusion speed of the radiation. The radiation is therefore trapped and advected inward.
The radius of the accretion disc increases steeply with the hole mass, until the whole 'plump' accretion disc grows 
in size enough to cross the trapping surface. At this point, an outflow develops, blowing away the disc.
A seed MBH can therefore grow in such fashion until $\sim 10^4 M_\odot$, bridging the gap between a massive seed and a 
supermassive hole.

In discs unstable to bar formation, instead, the direct formation of a seed BH and its rapid growth are indeed 
an appealing alternative (Begelman, Volonteri \& Rees 2006). 
The baryons in an unstable self-gravitating disc can lose angular momentum rapidly via 
runaway, global dynamical instabilities, the so-called ``bars within bars" mechanism. The accumulation of mass 
in the center leads to the formation of a dense core  which gradually contracts and is compressed further by subsequent infall, 
until the central temperature is so high that neutrino cooling takes over, leading to the formation of a seed black hole within the 
core (``quasistar'').  Subsequently, the black hole accretion rate adjusts so that the energy flux is limited to the Eddington limit 
for quasistar, rather than that of the seed BH itself, so the black hole grows at a super-Eddington rate as long as its mass is smaller 
than that of the ``quasistar''. The MBH could grow at a super-Eddington rate until it reaches $\sim 10^6 M_\odot$, at which point 
its mass would approach that of the ``quasistar''.  Thereafter it  could grow, by Eddington-limited accretion, to an even larger mass. 

Both the models for supercritical accretion apply only to metal-free halos with virial temperature  $T_{\rm vir} > 10^4$K, 
that is to rare massive halos, at the cosmic time considered. Rapid early growth, therefore, can happen only for a tiny fraction 
of MBH seeds, in a selective and biased way. These SMBHs are those powering the $z=6$ quasars, and later on to be found in the most 
biased and rarest halos today. The global MBH population, instead, evolves at a more quiet and slow pace.

\section{Conclusions}
The last few years have seen exciting developments in our understanding of the evolution of supermassive
black holes. Though we start having a coherent picture on how the population of SMBHs evolved, there are several 
crucial issues that remain to be clarified, both at high and low redshift.

%
%



\printindex
\end{document}